\documentclass{emulateapj}
\usepackage{amsbsy}
\usepackage{apjfonts}

\slugcomment{Accepted to Astrophysical Journal}
\shorttitle{TIME DEPENDENCE IN COLLISIONLESS SHOCKS}
\shortauthors{SPITKOVSKY \& ARONS}
\begin{document}
\title{Time-dependence in Relativistic Collisionless shocks: \\
Theory of the Variable ``Wisps'' in the Crab Nebula}
\author{Anatoly Spitkovsky\footnote{Chandra Fellow}} 
\affil{Kavli Institute for Particle Astrophysics and Cosmology, \\Stanford University, P.O. Box 20450, MS 29, Stanford, CA 94309}
\author{Jonathan Arons}
\affil{Astronomy Department, Physics Department and
Theoretical Astrophysics Center,
University of California, Berkeley, CA 94720}

\begin{abstract}

We describe results from time-dependent numerical modeling of the collisionless
reverse shock terminating the pulsar wind in the Crab Nebula. We treat the
upstream relativistic wind as composed of ions and electron-positron plasma
embedded in a toroidal magnetic field, flowing radially outward from the pulsar
in a sector around the rotational equator. The relativistic cyclotron
instability of the ion gyrational orbit downstream of the leading shock in the
electron-positron pairs launches outward propagating magnetosonic waves. Because of the fresh supply of ions crossing the shock, this time-dependent
process achieves a
limit-cycle, in which the waves are launched with periodicity on the order of
the ion Larmor time. Compressions in the magnetic field and pair density
associated with these waves, as well as their propagation speed,
semi-quantitatively reproduce the behavior of the wisp and ring features
described in recent observations obtained using the {\it Hubble Space Telescope} and
the {\it Chandra X-Ray Observatory}.  By selecting the parameters of the ion orbits
to fit the spatial separation of the wisps, we predict the period of time
variability of the wisps that is consistent with the data.
When coupled with a mechanism for
non-thermal acceleration of the pairs,
the compressions in the magnetic field and plasma density associated with
the optical wisp structure naturally account for the
location of X-ray features in the Crab. We also discuss the origin of the 
high energy ions and their acceleration in the equatorial current sheet of 
the pulsar wind.

\end{abstract}


\section{Introduction}

The mechanisms through which compact objects excite surrounding synchrotron
nebulae are one of the long-standing problems of high-energy and relativistic
astrophysics. The loss of rotational energy from the central magnetized object
underlies the synchrotron emission observed from pulsar wind nebulae (PWNs; e.g., Bandiera, Amato, \& Woltjer 1998) and is one of the prime candidates
for the energization of
the jets in active galactic nuclei, radio galaxies (see
Krolik 1999), and gamma ray bursts (e.g., Blandford 2002).
PWNs form the nearest at hand laboratories for the investigation of the physics
of such high-energy particle acceleration, and are the example where
rotational energization clearly
bears responsibility for the nonthermal activity.

Despite intensive study, however, the physics of how the rotational energy
gets converted to the observed synchrotron emission has not been
satisfactorily understood. Magnetohydrodynamic wind models (Kennel and
Coroniti 1984a,b and many subsequent efforts) provide an adequate
macroscopic description of the ``bubbles'' of synchrotron-emitting particles
and fields surrounding young ($t \sim 10^3$ yr) rotation-powered pulsars,
but identification and quantitative modeling of the mechanisms
that convert the unseen energy flowing out from the central pulsar
into the visible synchrotron emission have remained elusive.

Most attention has focused on the idea that the outflow is a supersonic and
super-Alfvenic wind, perhaps with embedded large-amplitude electromagnetic
waves (Rees and Gunn 1974; Melatos and Melrose 1996). Kennel and Coroniti's
model simplified the picture by assuming the wind to be steady, with no wave
structure. In that case, most models attribute the conversion of relativistic
outflow energy to the nonthermal particle spectra that emit the observed
synchrotron radiation in the body of the PWN to a relativistic
collisionless shock wave that terminates the flow. This idea was originated by
Rees and Gunn, who also observed that the shock should form at a radius where the
dynamic pressure of the outflow is approximately equal to the pressure of the
surrounding nebula.  They also noted that in the case of the Crab Nebula,
this radius approximately corresponds to the location of the moving ``wisps''
discovered 80 years ago (Lampland 1921) in optical observations of that
PWN. This identification has motivated many studies of the wisps, as possible
direct manifestations of the shock wave thought to underlie the transformation
of outflow energy into the nonthermally emitting particles in the nebula.

Shortly after the discovery of the Crab pulsar, Scargle (1969) described these
motions in more detail and attempted to interpret them as wave phenomena in 
the nebula stimulated directly by glitches in the pulsar's rotational spin-down
(Scargle \& Harlan 1970; Scargle \& Pacini 1971). Subsequent observations
made clear that the association of wisp activity to pulsar rotational activity
was a spurious consequence of undersampling the variability of the wisp
motions.

The advent of the magnetohydrodynamic model has focused attention on the
wisps' structure and variability as a probe of the physics of the region where
the pulsar's outflow interacts with the nebular plasma.  Substantially improved
observations in the optical (Hester et al. 1995; Tanvir, Thomson, \& Tsikarishvili 1997; 
Hester 1998a; Hester et al.,
2002), radio (Bietenholtz, Frail, \& Hester 2001) and X-ray (Mori
et al. 2002; Hester et al. 2002) now make it possible to
test relatively
refined theories of the shock's structure and particle acceleration properties.

A number of ideas have been suggested for the physics behind the observed
dynamic behavior.  Hester (1998a) suggested that the wisp structures reflect
synchrotron cooling at constant pressure of the particles whose outflow momenta
randomize at a shock in a magnetized electron-positron plasma to a power-law
distribution of particles with slope in energy space flatter than $E^{-2}$, an
idea that requires the X-ray spectrum of the plasma in the wisp region to be
flatter than $\varepsilon^{-1.5}$ (X-ray flux in
photons/keV-cm$^2$-sec). However, only the particles with energy high
enough to emit nebular gamma rays
($\varepsilon > 10 $MeV)
have synchrotron loss times short enough to create a cooling instability whose
time scales are comparable to the observed variations in the optical and X-ray,
and the gamma ray spectrum of the Crab (presumably arising in a compact region
around the pulsar no larger than the X-ray source) has a spectral index steeper
than 1.5.
Chedia et al. (1997) suggest that a drift instability of a subsonically
expanding plasma (no shock wave) can explain the wisp structure --
how the outflow from the pulsar
escapes catastrophic adiabatic losses is not addressed, however.
Begelman (1999) suggests that the wisps represent the nonlinear phase of a
Kelvin-Helmholtz instability between subsonic layers expanding in the
rotational equator of the pulsar
and flow at higher absolute latitude moving with a different velocity.

\begin{figure*}[t]
\centering
\plotone{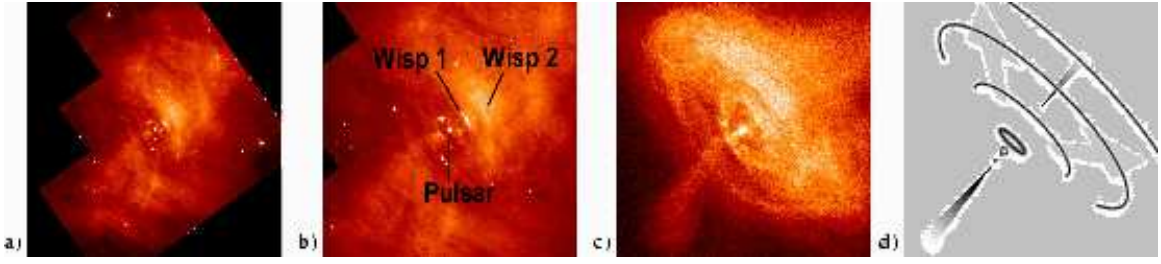}
\caption{(a) HST optical image of the inner Crab Nebula (after Hester 1995),
(b) Zoom in on the wisp region.
(c) X-ray nebula from {\it Chandra} (after Weisskopf 2000; [b] and [c] are
nearly to scale);
(d) Illustration of the polar jet and the equatorial flow geometry of the Crab
wind. {\it This figure is also available in color
in the electronic edition of the {\it Astrophysical Journal}}}.
\label{figwisps}
\end{figure*}

All of these ideas follow Kennel and Coroniti in assuming that the pulsar's outflow
decelerates in a shock wave of unobservably small thickness, with the wisp
dynamics due to phenomena in the post shock plasma (or, in the case of the
Chedia et al. model, do without a shock completely).  In contrast,
Gallant and Arons (1994, hereafter called GA), building on earlier work on the
structure and particle acceleration characteristics of relativistic shock waves
in collisionless, magnetized electron-positron (Langdon, Arons, \& Max1988; Gallant
et al. 1992) and electron-positron-ion (Hoshino et al.,  1992)
plasmas, attributed the observed wisp structure to the {\it internal} structure
of an electron-positron-ion shock wave. The idea is motivated by the fact that
the gyration of the heavy ions within the shock transfers energy to positrons
and electrons through emission of ion cyclotron waves (formally, the
magnetosonic mode of the magnetized electron-positron plasma). This leads to
the formation of nonthermal particle spectra in rough agreement with the
particle spectra required to explain the optical, X-ray, and gamma-ray emission,
if the ions carry a large fraction of the pulsar's total energy loss, at least
in the equatorial sector illustrated in Figure \ref{figwisps}. The
deflection of
the ions' outflow by the magnetic field deposits a large amount of
compressional momentum in the magnetic field, which is transverse to the flow,
and into the $e^\pm$ plasma frozen to the field.  These compressions appear as
brightenings in the surface brightness of the nebula.  GA showed that their
steady flow model gives a good account of a single high-resolution {\it I}-band image
of the wisps (van den Bergh \& Pritchett 1989).
However, the wisps are known to be highly variable, on a time-scale of weeks to
months.  An adequate account of their behavior requires a time-dependent
theory.

In this paper we present a model for time-variability
of the wisp region in the Crab by extending the work of GA to include
the time dependence in the flow. As was realized early on by Hoshino
et al.
(1992) and Gallant et al. (1992), the coherent ion orbits invoked in the
model of GA are unstable to gyrophase bunching and emission of magnetosonic
waves. We find that this instability introduces a time dependence into the
shock in the pair-ion plasma that causes the variable surface brightness to
have substantial similarity to the observed variability of the wisps.
Because of the presence of the continuous stream of fresh ions passing through
the pair shock, the relativistic ion cyclotron instability
provides a mechanism for sustained periodicity and wave emission dynamics
that closely reproduces time-resolved observations of the wisps.

In \S \ref{sec2} we describe the model
and the underlying assumptions. In \S\S \ref{shockdyn} and \ref{applic}, we
present the results of the simulations and the dynamical predictions for the
wisp region and compare them to observations.  In \S \ref{disc}, we discuss
the model's successes, with emphasis on its applicability to other PWNs that
have wind-nebula dynamics not very affected by radiation losses,
as well as the model's limitations, pointing out further
improvements needed in order to make fully quantitative comparisons with
observations. We also discuss some observational tests of the ideas at their
current level of development in this section. In \S \ref{origin} we describe
a scenario for the origin of the high-energy ions and their
acceleration in the equatorial region of the pulsar wind. Our
conclusions are summarized in \S \ref{conclu}.  The technical details of our
hybrid numerical approach are left for the Appendix.

\section{The model }
\label{sec2}
\label{model}


Following GA, we assume that the equatorial wind of the Crab pulsar
consists of relativistic electron-positron pairs and an ionic component,
which is minor in number density but energetically significant.
Because the ion gyration timescale is much larger than that of the pairs,
we resort to a hybrid model of the termination shock in the wind. In it, the
ions are treated kinetically, while the pair component is modeled as an
MHD fluid. When the magnetization of the wind is small,
the system is nearly charge and current neutral (see Appendix for
derivation; eq. (\ref{chderiv}-\ref{curderiv})).
Therefore, the charge and current density in the pairs can be found
by following the ion dynamics.

We assume the flow to be spherically symmetric in a $20^\circ$
sector about the equatorial plane. Using the spherical coordinates
($r$, $\theta$, $\phi$) centered on the pulsar, and allowing for variation
of all quantities with $r$ only, the conservation laws for mass, momentum,
    energy, and the field evolution equations can be written as

\begin{eqnarray}
&&{\partial \over \partial t}(\rho \gamma)+{1\over r^2} {\partial \over
\partial r}(r^2 \rho \gamma \beta_r)=0, \label{pcontinuity2}\\
&&{\partial \over \partial t}(\rho h \gamma^2 \beta_r
)+
{1\over r^2} {\partial \over \partial r}
(r^2
\rho h \gamma^2 \beta_r^2)
+ {\partial \over \partial r}P_\bot+{P_\bot
-P_\parallel\over r}
 \nonumber \\
&&\quad\quad-{1\over r} \rho h \gamma^2 \beta_\theta^2 + \epsilon N_i (\beta_{i\theta}
-\beta_{\theta})B_\phi=0, \label{rmomentum2}\\
&&{\partial \over \partial t}(\rho h \gamma^2 \beta_{\theta})+
{1 \over r^2}  {\partial \over \partial r}(r^2
\rho h \gamma^2 \beta_r \beta_{\theta} )
+{1\over r}(\rho h \gamma^2 \beta_r \beta_{\theta})
\nonumber \\
&&\quad\quad+\epsilon N_i (\beta_r - \beta_{i r})B_\phi
=0, \label{thmomentum2}\\
&&{\partial \over \partial t}(\rho h \gamma^2
-P_\bot)+
{1\over r^2} {\partial \over \partial r}(r^2
    \rho h \gamma^2 \beta_r )
\nonumber \\
&&\quad\quad+\epsilon N_i (\beta_{i \theta} \beta_r -\beta_{i r} \beta_\theta)B_\phi=0,
    \label{energy2} \\
&&{\partial \over \partial t} B_{\phi} + {1\over r^2} {\partial \over
\partial r}(r^2 B_\phi \beta_r)-{B_\phi \beta_r \over r}=0
\label{induction2}
\end{eqnarray}
where $\rho$ and $h$ are the pair density and specific enthalpy in
the proper frame,
$\gamma {\boldsymbol \beta}$ is the 4-velocity of the pairs,
$N_i=n_i \gamma_i$ is the observer's frame density of the ions,
and $B_\phi$ is the toroidal component of magnetic field, which is the only
magnetic field component that we consider in the equatorial plane.
$P_\bot$ and $P_\parallel$ are the components of the pair plasma's
pressure tensor perpendicular and parallel to $B$.
The field is normalized by the upstream value $B_1$, and expression
(\ref{sigmas}) defines $\epsilon$.
Since we are interested in the dynamics on the ion gyration
scale, we normalize time by $\omega_{ci1}^{-1}$ and radius
by $r_{Li1}$, where $\omega_{ci1} =Z e B_1/\gamma_{i1} m_i c$ and
$r_{Li1}=c /\omega_{ci1}$ are the upstream values of the
relativistic cyclotron frequency and ion Larmor radius, respectively, and $Z e$ is
the ion charge.
We normalize the enthalpy and pressure
by the upstream kinetic energy of the pairs
$m_\pm c^2 N_{1\pm } \gamma_{1}$, where $N_{ 1\pm}=\gamma_1 n_{1\pm }$ is
the pair number density in the observer's frame. After traversing the shock,
the pairs are confined to the plane perpendicular to the toroidal
magnetic field until pitch-angle scattering isotropizes the pair
distribution function.
We allow for the anisotropic pressure in the directions perpendicular
and parallel to the magnetic field,
$P_\bot \neq P_\parallel$.
These components of the pressure tensor are related through
the adiabatic index $\Gamma$ such that
$P_\bot=(\Gamma-1) \varepsilon$, where
$\varepsilon=2 P_\bot + P_\parallel$ is the internal thermal energy
of the pairs.
This results in $\Gamma=(3+P_\parallel/P_\bot) / (2+P_\parallel/P_\bot)$
with $\Gamma$ decreasing from $3/2$ to $4/3$ as pressure isotropizes.
In our model we simulate the effect of pitch-angle scattering in the pairs by
gradually decreasing  $\Gamma$ with distance from the pair shock over
a distance of $0.05$ pc (for further discussion of theory and simulation of pitch-angle
scattering in magnetized electron-positron-ion plasmas see Yang et al. 1993; Yang,
Arons, \& Langdon 1994).
The relative fractions of the preshock wind energy carried by the
ions, the pairs and the Poynting flux are parameterized using
\begin{eqnarray}
&&\sigma_\pm\equiv {B_1^2/4 \pi \over N_{1\pm}m_\pm \gamma_1 c^2}, \quad
\sigma_i\equiv {B_1^2/4 \pi \over N_{1i}m_i \gamma_{i1} c^2}, \quad \nonumber \\ 
&&\quad \quad \epsilon\equiv {\sigma_\pm \over \sigma_i}={N_{1i}m_i \gamma_{i1} c^2 \over
N_{1\pm}m_\pm \gamma_1 c^2}, \label{sigmas}
\end{eqnarray}
where $\sigma_\pm$ and $\sigma_i$ are the initial
magnetization of the pairs and ions, respectively. Global models of
the nebula (e.g.,
Kennel and Coroniti 1984a) constrain the total magnetization
$\sigma_{\rm tot}\equiv(1/\sigma_\pm+1/\sigma_i)^{-1}$ to be small,
$\sim 0.003$. Because of the small initial magnetization
of the wind and our attention to the region near the termination shock,
we neglect the terms proportional to $\sigma$ that otherwise would
enter into (\ref{pcontinuity2}-\ref{energy2})
(see Appendix for the full equations).

Equations (\ref{pcontinuity2}-\ref{energy2}) describe the system of
evolution equations that we solve numerically to find the time-variability
of the collisionless shock structure.
We use the particle-in-cell (PIC)
method (Birdsall and Langdon 1991) for advancing ions and calculating
their current and charge density, while the pairs are evolved by
finite-differencing the MHD equations on a one-dimensional mesh. Particular attention
is paid to the boundary conditions, with an outflow nonreflecting condition
(Thompson 1987) implemented on the downstream boundary.

\begin{figure*}
\centering
\plotone{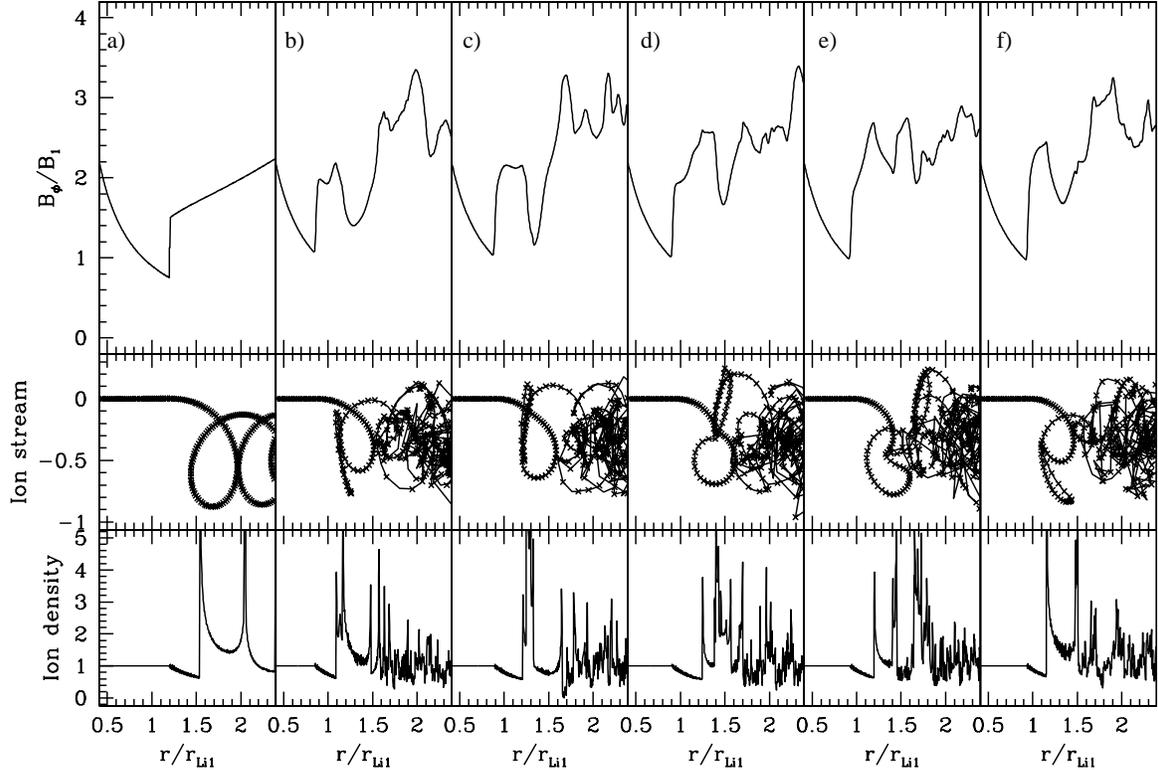}
\caption{Phases of the limit cycle. {\it Top}: Magnetic field, in units of
preshock value $B_1$, vs. distance. {\it Middle}: Ions' momenta
through the shock, with ion momentum in units of the upstream
momentum vs. distance, with length units of preshock ion Larmor
radii. {\it Bottom}: Ion
density in the radial direction, with density normalized to the
preshock ion density. The unit of length is the ion Larmor radius
based on upstream parameters, $r_{Li1}$. The columns are as follows: (a)
Initial state of the $\Gamma=3/2$ shock with $\Gamma$ decreasing to $4/3$ by
$r=1.5 r_{Li1}$. The cold ion stream is loaded through the pair shock
without affecting the field structure. (b-f) Developed
limit cycle. A gyroknot advects around the ion ring and triggers
the next cycle upon leaving the region. This figure is also available
as an mpeg animation in the electronic edition of the
{\it Astrophysical Journal}.}
\label{fig2cycle}
\end{figure*}

The
inner boundary condition for the pair plasma and magnetic field comes from the
jump conditions for an infinitesimally thin shock wave in the pair plasma,
as in Kennel and Coroniti (1984a). The shock is located inside the domain
and the flow is extrapolated upstream to the injection boundary at
$0.1$ $r_{Li}$. The ions enter as a cold stream
at the inner boundary, all with identical energy $\gamma_1 m_i c^2$,
where $\gamma_1$ is the Lorentz factor of the upstream wind --
we assume that the upstream ions and pairs flow with the same 4-velocity.
The jump conditions at the pair shock are such that
the magnetic field increases by a factor of $1/(\Gamma-1)$ over its
value in the upstream wind, and the velocity
of pair flow decreases by the same factor, while the
tangential electric field is continuous through the shock.
The ion stream, therefore, starts to gyrate as well as
${\bf E}\times {\bf B}$ drift in the magnetized pairs as
shown in Fig. \ref{fig2cycle}a.
The turning points in the ion orbits are regions of strong vertical
current that, through  the ${\bf j}\times {\bf B}$ force, provide a
compression on the
pairs. This compression results in locally higher density of the
pairs at each ion turning point, and compression of the frozen-in
magnetic fields. This field, in
turn, affects the motion of the ions, and hence the initial configuration is
bound to change. Although it is possible to load the ions and the fields
in a self-consistent static configuration, we find that the result of the
dynamical simulation is independent of the initial load method. We describe
the dynamical simulations in the next section.

\section{Shock dynamics}
\label{shockdyn}
The coherent ion loops generated during ion transit through the pair shock
are an excellent environment for the growth of a relativistic ion cyclotron
instability. This is a variant of the classic instability of a
``ring in momentum space'' and proceeds via gyrophase bunching and
subsequent phase mixing.
The crucial distinction in the shock case from the development of the
instability in a uniform plasma is that there is a preferred
direction of the flow, so that the time development of the instability
is spread out in space as the ions undergo $\bf{E}\times\bf{B}$ drift
into the shocked pairs.
The typical growth rate is on the order of the ion Larmor time (Hoshino
\& Arons 1991)
in the postshock field, and within the first ion loop gyrobunching develops
characteristic ``knots,'' shown in Figure \ref{fig2cycle}b. Beyond several
orbits in the postshock field the instability has taken its course, and the
downstream ion orbits become chaotic, with the ion distribution
eventually relaxing to a quasi-Maxwellian form. However, in
the case of a shock in a relativistic flow and in our simulation,
the cold ions are continuously flowing through the shock, thus constantly
refreshing the instability's flagging spirits.

The interplay between the low entropy input of ions
and the randomizing cyclotron instability creates interesting dynamical
consequences for the region of the first ion gyration loop.
In this region we observe the development of a limit cycle, in which
the first ion loop develops gyrobunched ion knots that are advected
downstream with the overall $\bf{E}\times\bf{B}$ flow. As one knot is
leaving this region, it creates an enhanced magnetic field that
triggers the development of a new ``gyroknot,'' which
goes around the ion ring and the process repeats. The snapshots of the
phases in this limit cycle are shown in Fig. \ref{fig2cycle}b-f.

In these frames we display the magnetic field, the ion stream
flowing through the
shock, and the corresponding ion density. The postshock
magnetic field shows the
characteristic ``double-horn'' structure due to compressions in the pair
density at the turning points of the zeroth-order ion drift orbit.
The development of the cyclotron instability in the first ion loop introduces
additional compression in the field because the gyrobunches in the ion stream
exert additional ${\bf j}\times {\bf B}$ stress on the pairs. As the gyroknot
is advected around the ion loop, the associated compression in the pairs
launches magnetosonic waves.
The area of the first ion ring  can be thought of as the near zone of
a radiating region, while the
flow further downstream represents the wave zone.
As the wave propagates between the turning
points in the first ion loop, it imprints a moving component
in the double-horn structure of the magnetic field. When this compression
in the field reaches the second ion turning point, it triggers gyrobunching
in the incoming ions. The period of the cycle is
found from simulations to be $\sim 0.5 T_{Li}$, where $T_{Li}= 2 
\pi/\omega_{Li}$ is the Larmor
time for the ions in the {\it postshock} field. The cycle period can be 
understood as roughly
the time in the laboratory frame that it takes for the ion to go around one loop in
the ${\boldsymbol E}\times {\boldsymbol B}$ drift orbit, or
$[(1-\beta_d)/(1+\beta_d)]^{1/2} T_{Li}$, where  $\beta_d c = (\Gamma-1) c$
is the initial drift speed in the downstream region.

\begin{figure*}[hbt]
\centering
\plotone{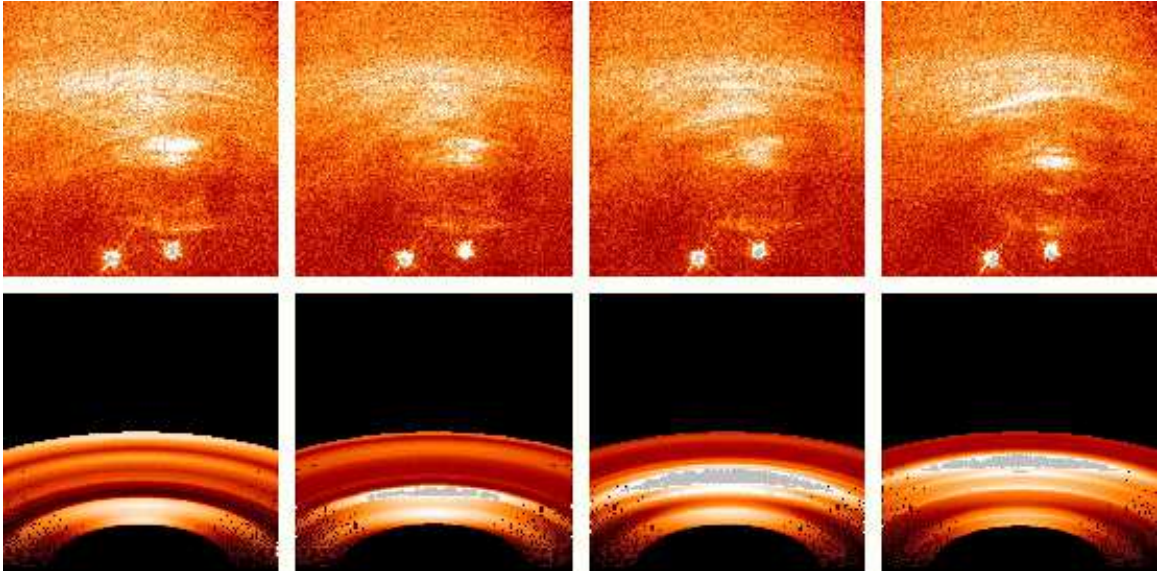}
\label{figsimul}
\caption{{\it Top}:
{\it HST} snapshots of the wisp region in the Crab. Separation between the
first and the last pair of frames is 3 weeks; middle frames are
separated by a week. The observation interval is from 1995 December 29 to 1996 February 22.
Data were extracted from the HST archive (observations by J. Hester).
{\it Bottom}:
Simulated surface brightness of the wisp region.
The moving wisp is being emitted from the first wisp and propagates toward
the second wisp. Fine filamentary structure is created in the second wisp region. 
This figure is also available in color and as mpeg animation
in the electronic edition of the {\it Astrophysical Journal}.}
\end{figure*}


\section{Application to the Wisps' Dynamics}
\label{applic}
As did GA, we interpret the wisp region in the Crab as a
collisionless shock in the pair plus ion plasma. Although
the ions are not directly observed, the compressions in the
pair density and magnetic field associated with the turning points in the
ion orbit should result in regions of increased synchrotron brightness.
GA associated the first two ion turning points with wisp 1 and wisp 2 (see
Fig. \ref{figwisps}b). Our time-dependent analysis augments this interpretation
with additional features.
We find that the back-and-forth motion of the gyro-knot
    in the region of the first
ion ring acts as a periodic source of magnetosonic waves that make the
shock structure very dynamic.
The magnetosonic waves that propagate back toward the pair shock
are reflected from it and compress the magnetic field near the shock,
brightening wisp 1. As the waves propagate from wisp 1 to wisp 2, they
appear as a moving bright feature, a ``moving wisp,'' that brightens
the region of wisp 2 when it crosses it.
The sequence of propagation and brightening of the features
is choreographed by the underlying limit
cycle of the instability in the first ion loop.
The overall timescale of the limit cycle can be obtained by choosing
the preshock ion Larmor radius $r_{Li1}$ such as to fit the observed
separation of the wisps 1 and 2 with the separation of compressions
in the model. The limit cycle then occurs with the period $T_{cycle}\approx
{1\over 2} T_{Li}=\pi (\Gamma-1) r_{Li1}/c$. Here we used the magnetic field
jump conditions to express the postshock Larmor time in terms of preshock
quantities. For typical values
$\Gamma=3/2$, $r_{Li1}=0.1$ pc we get $T_{cycle} \approx 6$ months.

We have taken our one-dimensional profiles of the magnetic fields,
    pair density and pair velocity to construct the sequence of
simulated surface brightness maps of the wisp region in the optical band.
We assume that in the proper frame the pair flow emits synchrotron radiation
as a relativistic Maxwellian. This is essentially a monoenergetic emission
and represents the surface brightness distribution in space for any
relatively narrow
bandpass from pairs whose Larmor radii are small compared to the ion
Larmor radius
(photon energies below 10 MeV). The radiation from the flow moving
along the equatorial
plane, tilted by $\sim 65^\circ$ from the plane of the sky, is further Doppler
boosted and beamed on the way to the observer.
The resulting maps, shown
in Fig. {\ref{figsimul}}, are a qualitative picture of the way the
features in the
wisp region should appear. The Doppler beaming makes the faster flow near the
shock appear to have smaller angular extent than slower flow in the region of
wisp 2. For comparison, we also reproduce snapshots from 1995-96 {\it Hubble Space 
Telescope (HST)} observations
of the inner nebula during
a moving feature emission episode in the same figure (Hester  et
al. 1998b). Observations from the 2000-2001 {\it HST} campaign (Hester 
et al. 2002) show
very similar dynamics of the features.
In principle, the simulated brightness profile can be constructed for
the X-ray band as well and will look very similar to the optical image. This
procedure will, however, overpredict the amount of beaming and make the inner
{\it Chandra} ring appear as two disjoint arcs, contrary to observations. We comment
further on this discrepancy in \S \ref{failure}.

Using the simulated surface brightness maps we can describe
the time dependent behavior of the wisp region throughout the
cycle of emission according to our model:

\begin{itemize}
\item[1)] The emission of a major moving wisp is preceded by a brightening
and thickening of wisp 1. This may also appear as a forward motion of wisp 1.
After the emission, it dims and retracts to the original position.
In general, the oscillating pressure of magnetosonic waves in the wisp region
causes oscillation in the position of the shock in the pairs. This
can make wisp 1 appear as oscillating with the cycles of
emission of the moving wisps.
The amplitude of this oscillation is relatively small ($<10\%$
of the distance between wisp 1 and the pulsar).

\item[2)]
The compression associated with the moving wisp increases in brightness
while propagating toward wisp 2. Because of this, the moving wisp
appears to turn on some distance away from wisp 1 ($<20 \%$ of the
distance between wisp 1 and wisp 2). The moving wisp appears as
a fairly sharp feature.

\item[3)]
The region of wisp 2 is located at the interface of the first and
second ion loops. By the second gyration loop the ion cyclotron instability
is well into the nonlinear stage, and the ion orbits are scrambled
and chaotic. This ion background provides a large amount of short-wavelength
features that comove with the flow as the ions drift downstream.
In fact, observations by
Hester et al. (1995) indicate that wisp 2 has a fine fibrous
structure. We attribute the creation of such filaments to the chaotic
ion background. Although the fine filaments should appear moving away
from the shock roughly with the flow velocity, we expect that new
filaments can appear in this region
without having a precursor traveling all the way from the shock.

\item[4)]
When the moving wisp propagates toward wisp 2, it brightens up wisp 2 and
its surroundings. The moving wisp is a compression in the field and pair
density that propagates as a fast magnetosonic wave.
This wave was  emitted from the first postshock ion
gyration loop and travels at the fast magnetosonic speed
on top of the background pair flow. In the case of low magnetization
the wave speed is given by $c(\beta_{\rm flow}+
\sqrt{\Gamma-1})/(1+\beta_{\rm flow} \sqrt{\Gamma-1})$. Assuming
that the pairs have isotropized
in the region of the second wisp, so that  $\Gamma=4/3$, we get
    $v_{\rm wisp}\approx
0.6$c as would be seen in the equatorial plane of the Crab.
When the moving wisp traverses wisp 2, it overtakes slower moving
fibrous filaments, and causes more filaments to be created in its wake, as
it scatters on the random ion background.

\item[5)]
The propagation of the fast-moving wisp from the first to the second wisp
takes about half the cycle of the wisp emission ($\sim 90$ days).
For the second half of the cycle no major brightening occurs, until the
first wisp starts to brighten again, indicating the beginning of the next
cycle. This behavior paradoxically results in the wisp structure
appearing both stationary and moving at the same time.
On average there are always
two bright regions corresponding to the first and second wisps, with the
second wisp being more diffuse. These regions correspond to compressions due
to the turning
points in the first ion loop, which are more or less fixed in space, but can
oscillate in brightness.
At the same time, the instability in the ion gyration produces
intermittent wisps that move between the first and second wisp at the
magnetosonic speed and change the brightness of the main wisps as they
compress the plasma and the field. Yet, the first and second wisps do not
move as a wave themselves.
\end{itemize}

The model has several parameters that we can vary to study the shock dynamics
in qualitatively different regimes. The main parameters are the ratio $f$ of
the preshock ion Larmor radius to the radius $r_s$ of the shock in
the pairs, $f\equiv r_{Li1}/r_s$,
and the relative fraction $\epsilon$ of the energy density carried by the ions
and the pairs (eq. [\ref{sigmas}]). The parameter $f$ controls the degree to
which spherical geometry affects the shock dynamics. For small values of $f$,
the background flow is essentially plane-parallel on the ion gyrational scale,
and the ions do not experience the gradient in the background magnetic field.
For the Crab parameters, $f\sim 1$, so the effects of spherical geometry are
important. Increasing $f$ decreases the separation of the turning points
in the ion orbit and changes the relative strength of plasma 
compressions corresponding
to the turning points, as illustrated in Figure 7 of GA.
The parameter $\epsilon$ controls the amount of influence the ions
have on the pair fluid. For small $\epsilon$ the underlying pair
shock structure
is hardly perturbed. This makes the instability grow at a slower rate, and the
first several ion loops may appear steady. The gyro-knots, characteristic of
the instability, then appear only several Larmor radii downstream. In the limit
$\epsilon\to 0$ there is, of course, no instability, and the ions and the pairs
are decoupled.
For finite $\epsilon$, once the instability
proceeds, the period of the limit cycle is not sensitive to 
$\epsilon$, and is approximately
half of the local Larmor time of the ions.
We fix the value of $f \approx 1.2$ to match the separation of
the wisps with the compressions at the turning points in the ion orbits. We
then vary $\epsilon$ to achieve the conditions where both the instability
grows sufficiently by the time the ions cross the first ion loop, and the
compressions in the magnetic field and the density result in the intensity
variations consistent with observed variation in the nebula.
For perturbations
$\delta B/B \sim 1$ we need $\epsilon \sim 1$. For the pair injection rate of
$\dot{N}_\pm\approx 10^{38}$ $\rm{s}^{-1}$ needed to explain the X-ray source
in the Crab, we then find that the ion injection rate should be
$\dot{N}_i=\epsilon m_\pm/m_i \dot{N}_\pm \approx 10^{34}$ $\rm{s}^{-1}$ if
we assume that the ions are helium
(the model favors ions with a $Z/A$ ratio of $\sim 2$, which could be
either He or Fe). These ions travel with the same Lorentz factor ($10^6$) as
the equatorial pair wind.

\section{Discussion }
\label{disc}
\subsection{Model Successes}

The theory outlined here has a number of virtues in its description of the
physics of the termination of the Crab pulsar's equatorial wind.  The timescale and
length scales of the observed optical and X-ray variations can be
semi-quantitatively reproduced, for entirely reasonable values of the ion
energy.  Granted the approximate validity of the underlying Kennel and Coroniti
(1984a) flow model, for ions that accelerate through a fraction
$\sim 0.1-0.2$ of the available
open field potential $\Phi\approx 4 \times 10^{16}$ V to energies
$\sim 0.15 Ze\Phi \approx 6 \times 10^{15}$
eV, the ion Larmor radius, which sets the wavelength scale of the moving
features, is $r_{Li} \sim 10^{18} B_{s16}^{-1}(r_s /r) $ cm. Here $r_s \leq
1.5 \times 10^{17}$ cm is the radius of the shock in the pairs, located at or
interior to the {\it Chandra} X-ray ring and $B_s = \sigma^{1/2} B_{nebula} \approx
16 B_{s16} \; \mu$G. The ion energy is obtained by asking the model to
replicate the spacing of the moving features in the optical and X-ray
images. The ions are deflected and become magnetically well coupled to the pair
plasma in the increasing magnetic field at $r > r_{turning} = 
[r_{Li}(r_s) r_s ]^{1/2}
\approx 4 \times 10^{17} (\gamma_{i6.5}/B_{s16})^{1/2}$ cm. This is the radius
at which the compressional limit cycle formed by the first ion loop stands as a
quasi-stationary feature in the pair flow.  Given the close resemblance between
this feature of the model and the behavior of the {\it Chandra} X-ray ring, it is
tempting and natural to identify the X-ray ring with the first ion loop's
turning points.  If so, the quasi-stationarity of this feature then
has a natural
explanation, as being the consequence of the ion cyclotron instability within
the shock structure, without having to invoke extra physics and an otherwise
unexplained external stimulus to account for the time scale of the ring's
repetitive variations.  The non-axisymmetric knotted structure of the ring
might be explainable as the consequence of instabilities of the mirror type,
driven by the reflected ions' pitch angle anisotropy with respect to the
toroidal magnetic field -- a topic well beyond the scope of the current
investigation, however.

The model also gives a natural interpretation of the moving features
in both optical
and X-ray, as compressional waves in the pair plasma emitted by the time
variable ion currents in the first loop limit cycle.
These modes are a mixture of magnetosonic waves moving with respect
to the plasma flow
and entropy and slow modes frozen into the flow. This behavior
supports the identification of
the {\it Chandra} ring as the location of the ion limit cycle structure,
since that ring's
appearance suggests it to be the wave emission zone.

The theory is adiabatic with respect to radiation losses from the pairs; that
is, the formation of the ring and the moving wisps does not depend on the
synchrotron loss times of the pairs being comparable to or shorter than the
local flow time in the vicinity of the ring. Since ringlike and wisp features
have been discovered near pulsars in other PWNs at radii where dynamic pressure
balance arguments suggest formation of wind termination shocks, yet all occur
in systems whose synchrotron cooling times even for X-ray-emitting particles
are larger than the local flow times, the theory developed here appears to
have general applicability, in contrast to models that depend on the unusually
strong magnetic fields and short synchrotron cooling times prevalent in the
Crab Nebula (e.g., Hester 1998a; Hester et al. 2002).
Gaensler et al.
(2002) give an example of our theory's use, in the interpretation of the X-ray
features observed close to PSR 1509-58.

This model has a substantial theoretical virtue: the required ion
flow has the magnitude expected if it is the electric return current
(Goldreich-Julian current) that should exist in the rotational equator of the
pulsar's wind (Michel 1974; Contopoulos et al. 1999), which
prevents the
star from charging up when non-vacuum torques are a significant contributor to
spindown.  This conclusion is the same as was reached by GA from a
time-independent flow model, now seen as a robust consequence of the fully 
time-dependent theory.

\subsection{Model Limitations and Failures }
\label{failure}
There is a substantial ($\sim $20\%) discrepancy between the velocities of
the moving magnetosonic wave features computed in the model ($0.6c$) and
the speeds of the observed features [$\sim 0.5c$, in the geometry adopted by
Hester et al. (2002), itself a velocity large compared to what one would
expect, $(c/3)(R_{shock}/r)^2$, for features frozen into a simple Kennel \&
Coroniti 1984a flow].  This discrepancy may be related to the geometrical
simplifications of the model, discussed further below, or it may be an
intrinsic difficulty.

The model's one-dimensionality is a serious limitation.  We assumed the flow to
be strictly radial, within a spherical sector around the rotational equator.
Within that sector, we assumed the magnetic field to be strictly toroidal and
have a single sense of winding with respect to the rotation axis of the pulsar,
independent of rotational latitude.  Going along with this 
one-dimensionalization, we assumed the flow of pairs and ions to be completely
charge neutral and to be current neutral in the radial direction.
Yet we concluded that the ion flux corresponds in
magnitude to an electric current flowing out in the equatorial plane with
strength sufficient to qualitatively and quantitatively alter the magnetic
field.

A more realistic model would take this result seriously and treat the ion flux
as a net electric current in the radial direction (balanced in a quasi-steady
state by a polar electron current feeding the polar jets). Such a current in
the upstream wind separates two hemispheres of toroidal (and very weak radial)
magnetic field that are oppositely directed in the two hemispheres.  In such a
field, high-energy ions in the equatorial current flow, upon crossing the
shock in the pairs, are not fully reflected by the compressed magnetic field in
the heated pair plasma. They are deflected from their initial radial orbits,
but transfer a smaller fraction of their outward momentum into compressions of
the pair plasma. Thus, the model described here somewhat overestimates the
compression of the magnetic field in the region where the limit cycle appears,
probably by a factor on the order of 2.  Generating synthetic surface
brightness maps, as a function of assumed ion and pair energy and number
fluxes, that are quantitatively reliable requires taking this more realistic
geometry and ion dynamics into account.

Several comments are in order regarding the speed of propagation of the wisps.
Quasi-periodic temporal drive due to the ion instability leads us to interpret
the moving features in the Crab as the fast MHD waves. This, however, leads not
only to overestimation of the speed of the features as outlined above, but also
to an uncomfortable prediction that the wisps should reach an asymptotic
velocity as they propagate downstream. As the background flow decelerates, the
speed of the disturbance decreases to the sound speed $\sqrt{\Gamma-1}c=0.58c$
for $\Gamma=4/3$.  Fast waves are not the only type of perturbations generated
by the instability, however.  Ion motion also excites entropy, Alfven and slow
waves. All of these perturbations have zero phase velocity for the toroidal
background magnetic field; therefore, they are advected with the
flow. The entropy wave does not have any perturbation in the magnetic
field and
therefore will not appear as bright as the modes that perturb both the density
and the magnetic field.  The Alfven mode is peculiar because it is the only
disturbance that has the transverse velocity perturbations.  It can
be advected with
the ion flow that generates the perturbation in the pairs' latitudinal
velocity.

As the fast wave from the emission region crosses wisp 2 and
encounters the chaotic ion background beyond the first ion ${\boldsymbol
E}\times {\boldsymbol B}$ loop,
there will be some compressions in the pairs and the field that move with the
average ${\boldsymbol E}\times {\boldsymbol B}$ velocity of the ions
(which is the same as the pair flow
velocity). These compressions are not free-propagating, but driven by
${\boldsymbol j}\times {\boldsymbol B}$ forces from ion streams.
They are similar to the
  double-horn soliton structure generated by ion loops (Alsop \& Arons 
1988), but
  they are moving as the ion loops are advected with the unstable ion stream.
Therefore, at larger distances it is reasonable to expect
wisplets that appear to move with the background flow. However,
such features would be
relatively short lived (approximately a month) and would appear to come seemingly from
nowhere. Therefore, beyond wisp 2 we expect to see two populations of
wisps -- ones
moving with the background decelerating flow (slow magnetosonic and Alfven
waves), and fast magnetosonic perturbations that can be traced back to the
shock in the pairs at wisp 1. The intensity of the fast compressions will get
diminished with distance, and they will naturally blend into slower background
flow.  A clear observational tracking of a single feature through the whole
flow (with snapshots separated by no more than 2 weeks) would be very
beneficial to settling the issue of the speed of the flow and the
nature of the speed of the features.

The front-to-back asymmetry of the optical wisps' brightness has an obvious
interpretation in the Doppler boosting enforced by the underlying flow, which
preserves on average the Kennel and Coroniti radial profile
$v_r = (\Gamma -1) c (r/R_s)^2 $. However, the inner {\it Chandra} ring does not show
such obvious asymmetry, even though it appears to be spatially colocated
with wisp 1.  If this apparent symmetry is real, our identification of the first
ion turning point with the {\it Chandra} ring obviously conflicts with our
interpretation of
the optical wisps.  We believe that this discrepancy is another
manifestation of the one-dimensional model's limitations.  If
the {\it Chandra} ring forms in the equatorial current layer, which is embedded in
a higher latitude ($|\lambda | > 10^\circ $) flow (Begelman 1999),
perhaps having
higher $\sigma $ (Arons 2004), the flow right in the equator might have lower
velocity than the higher latitude component, with the optical wisps
being features
stimulated in the higher latitude flow by the large Larmor radius ions.
Such a multi-dimensional environment, which is implied by the ions being a
net electric current, then might account for Doppler boosting of the
optical wisps
along with smaller or negligible Doppler boosting of the X-ray
features. Testing
dynamical viability of such a scheme awaits the development of a
two-dimensional, axisymmetric model, whose MHD aspects are likely to be
similar to the nebular model recently proposed by Komissarov and 
Lyubarsky (2003).

\subsection{Nonthermal Particle Acceleration and Radiation \label{rad_accel}}

Although our dynamical model and synthetic images assume a thermal
(essentially, a monoenergetic) pair distribution in the radiating plasma, they
can be extended to the regime where the distribution is nonthermal, as is the
case for the X-ray and optically emitting particles. The compressions in the
field and plasma density associated with the shock structure should cause
higher synchrotron emission for all electrons and positrons with Larmor radii
smaller than the characteristic wisp separation, or for $\gamma_\pm <
10^{10}, \, E_\pm < 5 \times 10^{15}$ eV, emitting at $160 \; {\rm keV} <
\varepsilon_{max} < 50-100$ MeV in the magnetic field whose strength increases
from $\sim \sqrt{\sigma} B_{nebula} \approx 16 \, \mu$G at the leading edge of
the shock structure to $\sim 300 \, \mu$G at and beyond the outer reaches of
the X-ray torus (Kennel and Coroniti 1984a; de Jager and Harding 1992).  Thus,
the wave motions seen in the {\it Chandra} band should display behavior similar to
the optical wisps; that is, the features emitted from the X-ray inner ring
discovered by Weisskopf et al. (2000) display similar variability as the
optical moving wisps, and all propagate toward and through the X-ray
torus. The observations do show such resemblances, at the qualitative level.

We have identified the {\it Chandra} X-ray ring as the site of the limit cycle in the
unstable, reflected ion stream, not as the leading edge reverse shock in the
pairs. The main reason to do this is the morphological similarities between the
observed ring and our synthesized image. Whether the actual pair shock is
located at the {\it Chandra} ring as concluded by Hester et al. (2002) or
interior to it depends on the shock acceleration mechanism for the pairs.
Below we consider the consequences of several shock acceleration schemes.

The pair shock heats the pair plasma to characteristic particle energies $\sim
\gamma_{wind} m_\pm c^2 \approx 10^{12} \gamma_{w6.5}$ eV, where
$\gamma_{w6.5} = \gamma_{wind}/10^{6.5}$.  The downstream magnetic field at and
immediately behind the pair shock is weak.  For
$\sigma \approx 3 \times 10^{-3}$, $B \approx 16 B_{s16}(r/R_{s\pm})
\, \mu$G.  Pairs compressed and heated by the pair shock
radiate in the optical and infrared, with characteristic frequency
$\nu_c = 6 \times
10^{14} \gamma_{6.5}^2 (r/R_{s\pm})$ Hz.  However, the synchrotron radiation time
of newly shocked pairs at and just behind the pair shock is long, $T_{s\pm}
\approx (3800/B_{s16}^2 \gamma_{6.5}) (R_{s\pm}/r)^2$ yr, far larger than
the local flow time $T_{\rm flow} = r/v(r) = 0.5 (R_{s\pm}/ 10^{17.2} \, {\rm
cm})(r/R_{s\pm})^3$ yr.  Thus simple shock heating of the pairs implies
that optical emission does not start until $r > 10^{18}$ cm from the pulsar, which
is larger than the radius of optical emission onset.

Gallant et al. (1992) demonstrated that the pair shock can indeed
thermalize the upstream flow. The pair shock may, of course, also be a
nonthermal particle accelerator, as has been assumed in all MHD models for the
excitation of the Crab, starting with Rees and Gunn (1974).  Diffusive Fermi
acceleration (DFA) in transverse relativistic shocks is often invoked as the
mechanism for such acceleration (see Gallant 2002 for a review).  Since the
energy gains per encounter of a particle with the shock are $\Delta E \sim E$,
this process might turn the pair shock into a nonthermal X-ray emitter, if the
cross field diffusion time is short in the downstream medium, comparable to the
cyclotron time (``Bohm diffusion'') -- such diffusion must occur, if the
requisite multiple encounters are to exist.  Diffusion speeds exceeding the
downstream flow speed ($\sim c/3$) require quite strong downstream turbulence,
however (Bednarz and Ostrowski 1996), probably at levels higher than observed
in the two-dimensional pair shock simulations of Gallant et al. (1992).
In the absence of three-dimensional simulations that allow evaluation of the turbulence level
and that also allow for cross field diffusion, DFA at the pair shock is not
excluded as the means of creating nonthermal emission, although the constraints
on the required diffusion rates clearly are rather demanding.

Progressive acceleration with increasing radius, due to the resonant
cyclotron absorption by the pairs of the high harmonic ion cyclotron
waves emitted by the
ions (Hoshino et al. 1992; E. Amato \& J. Arons, in preparation), is an
alternate to the commonly assumed DFA that has a natural setting in the context
of the ion-doped relativistic shock whose spatial structure has been modeled
here.  The acceleration time is approximately the ion cyclotron time, which
leads to energy gains such that X-ray emission from the {\it Chandra} ring requires
placing the now unobserved pair shock at a radius between 2 and 3 times smaller
than the radius of the X-ray ring.  A full discussion of this distributed
acceleration model will be given elsewhere.

\subsection{The Acceleration of the High Energy Ions in the Wind}

\label{origin}

So far we have not addressed the issue of how the ions are energized to 
the inferred values of $\gamma \sim 10^6$.
One possibility is that the high energy of the ions and the high 
4-velocity of the pulsar wind are due to the gradual ``surf-riding'' 
acceleration in the inner wind of the pulsar. 
If
the wind begins near the light cylinder with $\sigma_{L0} \equiv 
\gamma_L \sigma
(R_L) \gg 1$, where $\gamma_L \equiv \gamma(R_L)$ is the Lorentz factor of a
particle injected into the wind at $r=R_L$, then 
particles riding on
the field lines of the wind move on almost radial paths and increase their
energies linearly with radius (Buckley 1977). This acceleration continues 
until the almost force-free, multidimensional flow in the inner region ends
at the radius $R_{ff}$, which is not greater than $\sigma_{L0}^{1/3} R_L$, 
where $\sigma (r)$ drops to
$\sigma_{L0}^{2/3}$ and $\gamma_{wind} \approx \sigma_{L0}^{1/3}$
(Beskin, Kuznetsova, \& Rafikov 1998; J. Arons 2004, in preparation). Here 
$R_{ff}$ is the radius
where the flow speed first reaches the velocity of the fast magnetosonic wave;
beyond this radius, the magnetic stress no longer can overcome the longitudinal
inertia $\rho \gamma_{wind}^3$ of the plasma, where $\rho$ is the 
rest mass density.
The acceleration in the inner region has its origin in the
particles being
stuck to the field lines, with the balance of electromagnetic forces
causing the field lines to move with an ${\boldsymbol E}
\times {\boldsymbol B}$ velocity
that is slightly below $c$, but
increasing toward $c$ with distance. This effect 
results in
$\gamma mc^2$ increasing linearly with radius for all particles in the wind,
when $r < R_{ff}$. For $r > R_{ff}$, $\gamma_{wind}$ stays 
approximately constant
in laminar, ideal MHD relativistic flow. This conclusion has been 
formally demonstrated
only when the poloidal field is monopolar, but is likely to be a 
robust conclusion
in practical terms from all relevant magnetic geometries (Begelman and Li 1994;
J. Arons 2004, in preparation).\footnote{Early one-dimensional MHD models
(Michel 1969; Goldreich \& Julian 1970) found $\gamma \rightarrow 
\sigma_{L0}^{1/3}$
as $r \rightarrow \infty$, a result that is an artifact of their 
one-dimensional
(in effect, cylindrical) flow assumptions.  The fact that $R_{ff} < \infty$ in
the multidimensional model of Beskin et al. (1998) results from their
incorporating the transverse (to the radial flow direction) electromagnetic
stresses, while their result that $R_{ff} =\sigma_{L0}^{1/3} R_L \gg R_L$ is an artifact of 
using a strict monopole for the poloidal magnetic field. More physical models
that have modest departures of the poloidal field from the monopole have
$R_{ff}$ at most a few times $R_L$. Contopoulos and Kazanas (2002) 
are incorrect
in their claim that the force free region extends to $\sim 
\sigma_{L0} R_L$ with
$\gamma \rightarrow \sigma_{L0}$ -- the acceleration dynamics is well described
by force-free theory in a much more limited region than they assumed.}

Such a linear accelerator in the inner wind, an idea recently resurrected
by Contopoulos and Kazanas (2002), reaches finite-inertia saturation too soon
to provide sufficient acceleration to the wind and the ions in the Crab.
Our results
for the particle content of the wind's equatorial sector with latitude range
$|\lambda | < 10^\circ $
suggest a Goldreich-Julian flux of heavy ions
$\dot{N}_i = {\dot{N}_{GJ}}/Z= 2\Omega^2\mu/Zec \approx 2 \times
10^{34}/Z $ s$^{-1}$, and a flux of
electron-positron pairs
$\dot{N}_+ + \dot{N}_- = 2 \kappa_\pm {\dot{N}_{GJ}} \approx 10^{4}
\dot{N}_i \sim 10^{38.3} $ s$^{-1}$ (Harding \& Muslimov 1998;
Hibschman \& Arons 2001;  Muslimov \& Harding 2003).
Then
\begin{eqnarray}
&&\sigma_{L0}  \equiv 
    \frac{B_L^2}{4\pi [m_i n_i(R_L) + 2 m_\pm n_\pm (R_L)] c^2}
     = \frac{1}{2} \frac{Ze\Phi}{m_i c^2}
        \frac{m_i}{m_{eff}} ,  \\
&&\quad \quad m_{eff}  \equiv  A m_p + 2 \kappa_\pm Z m_\pm . \label{eq:m_eff}
\end{eqnarray}
For Crab pulsar parameters (magnetic moment $\mu \sim 10^{30.5}$ cgs,
$\Omega \approx 200$ s$^{-1}$, $\Phi \approx 4 \times 10^{16}$ V) and
equatorial plasma properties inferred from our
model of the wisp and ring dynamics ($Z = 1$ or $2$, $m_i = A m_p$ with
$A = 1$ or $4$ and $2\kappa_\pm \sim 10^{4}$),
\begin{equation}
\sigma_{L0} \approx (1.3-1.7) \times 10^6  \gg 1.
\end{equation}
Our model for the wisps' dynamics suggests
$\gamma_{ion} \sim 0.1 Ze\Phi /m_ic^2 \approx \sigma_{L0} \gg 
\sigma_{L0}^{1/3}$.
Since laminar acceleration of the whole relativistic wind does not explain
the inferred $\gamma_{ion} \approx \gamma_{wind} \sim 10^6$ or
$\sigma_{wind} (r_s) \sim 10^{-2.5}$ in the equatorial flow (the only 
latitudes where
we have substantial quantitative estimates for $\gamma$ and $\sigma$ 
in the wind),
some additional physics must be at work to accelerate and demagnetize 
the wind in the
equatorial sector.

The fact that the wisps (and the X-ray torus) lie in an angular sector near the
pulsar's rotational equator and that these regions appear to accumulate the
majority of the instantaneous rotational energy loss offers an important clue
to the additional acceleration physics, a clue replicated in an increasing number of other
young PWNs. That clue suggests that the equatorial region of this
relativistic outflow has special properties not shared with the higher latitude
regions, perhaps arising from the electric (return) current flowing in a
crinkled sheet structure in an equatorial angular sector.
Substantial theoretical evidence has accumulated to suggest that the
$\sigma \gg 1$ regions
of pulsar winds have structure close to that of a split monopole for $r$ larger
than a few times
$R_L$.  This result is quite certain for the aligned rotator, both
from the asymptotic analysis
(Michel 1974) and from the self-consistent numerical
solution recently obtained by
Contopoulos, Kazanas and Fendt (1999). While there is no similar
demonstration that
the asymptotic structure of the force-free oblique rotator's wind
{\it must} be close to that of the
oblique split monopole, whose field structure was recently obtained
by Bogovalov (1999),
we regard Bogovalov's model to be an excellent representation of the
likely structure
of the electromagnetic fields in the wind of an oblique rotator
outside the light
cylinder and will assume those fields in this discussion of wind acceleration.

\begin{figure}
\plotone{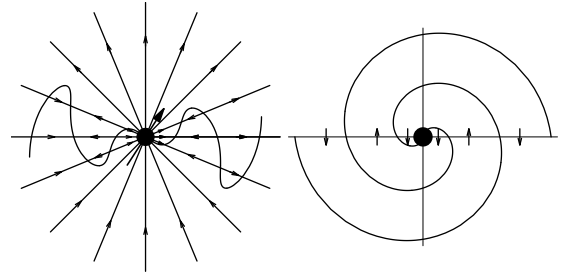}
\caption{Frozen-in current sheet structure of the oblique split monopole in the
inner wind.
{\it Left}: Meridional cross-section of the poloidal field structure,
showing the crinkled current sheet. The thick arrow indicates the 
instantaneous direction
of the magnetic axis. Rotation is around the vertical axis.
{\it Right}: Intersection of the current sheet with the equatorial plane.
The toroidal magnetic field forms stripes with opposite directions
between each current
layer, as indicated by the arrows between the current sheets.}
\label{fig:split_mon}
\end{figure}

In the equatorial sector $-i < \lambda < i$, with $i$ the inclination
angle between the magnetic moment and angular velocity vectors,
the electromagnetic fields have the form
\begin{eqnarray}
&&B_r  = \frac{M}{r^2}f\left[\frac{r}{\beta R_L} - (\phi - \Omega t)
\right],\nonumber \\ 
&& B_\phi  = -\frac{M \sin \theta}{rR_L}f\left[\frac{r}{\beta R_L} - (\phi -
\Omega t) \right], \; \nonumber \\ 
&& E_\theta = B_\phi, \quad   
B_\theta =  E_r = E_\phi = 0. \label{eq:fields} 
\end{eqnarray}
Here $r, \theta, \phi $ are the spherical radius, colatitude
(measured from a $z$-axis
with positive $z$ along the neutron star's angular velocity vector),
and rotational
azimuth, respectively, measured from an arbitrary $x$-axis with $\phi$ increasing
in the direction
of rotation. Furthermore, $f = \pm 1$ is a step function of pattern phase
$(r/\beta R_L) + \phi - \Omega t$, where $c\beta$ is the radial
outflow velocity.
The fields reverse in sign every wavelength [at $r=(n+1)\pi \beta R_L
/2$], while
the magnitudes of the fields remain the
same as those of the aligned split monopole. The effective monopole moment
is $M = k \mu/R_L$, with $k$ a constant on the order of unity; $k =1.36$ in the
Contopoulos et al. (1999) solution for the aligned rotator. A
crinkled current sheet
separates the regions of oppositely directed field, shown in Figure
\ref{fig:split_mon} (following Bogovalov 1999).
As is the case with the Deutsch (1955) fields of the vacuum rotator,
these fields form a spirally wound pattern that corotates with the star.
Outside the equatorial sector, $f =$ constant. For $i < \pi/2$, as is
thought to be the
case for the Crab pulsar (Yadigaroglu \& Romani 1995), $f = 1$ for
$0 \leq (\pi/2) - i$
and $ f= -1$ for $(\pi/2) + i \leq \theta \leq \pi $.

In the global, ideal model of the current flow, the ions reside in the
crinkled current
sheet, which is frozen into the flow.  Therefore,
they participate in the same acceleration to which the
plasma outside the current sheet is exposed. In the global wind models
the current sheet has infinitesimal thickness.  However, in reality the current
flows in a crinkled layer of finite thickness and may be subject to
dissipative processes that can lead to the release of magnetic energy in the
equatorial sector in the many decades in radius between $r = R_{ff} 
< 10^2 R_L$
and $r=r_s = 10^9 R_L$. The dissipation of magnetic energy can lead to
acceleration (and possibly heating) of the plasma
(including the ions) to higher 4-velocities than is possible in the 
ideal flows so
far studied.  Whether such dissipation is inductive destruction of the
current sheet, modeled as a sheet pinch (Coroniti 1990; Michel 1994;
Lyubarsky \& Kirk 2001, Kirk \& Skj{\ae}rassen 2003), or has
something to do with the dissipation of relativistically
strong electromagnetic waves (Melatos 1998), perhaps generated by instability
of the current sheet and accelerating the ions through ponderomotive wave
pressure (Arons 2003, 2004), remains an open question.

\section{Conclusions }
\label{conclu}
We have studied the internal structure and time variability of a collisionless
shock in an electron-positron pair plasma with an energetically significant
admixture of ions. Upon crossing the shock the ion component undergoes
relativistic cyclotron instability that renders the shock structure very
dynamic. The first loop of the ${\bf E}\times {\bf B}$ ion orbit
achieves a limit-cycle
behavior and acts as a quasi-periodic wave emitter with a period of half a
local Larmor time of the ions. This time-dependence led us to apply our
results to the collisionless termination shock in the inner Crab Nebula where
we identify the wisps with turning points in the drift orbit of the ions. The
limit cycle of the instability produces large amplitude magnetosonic waves on a
time-scale of several months that proceed to move downstream from the
shock. This time-scale is intrinsic to the model and associates the
appearance of the wisps to the mechanism of variability within the shock
structure.  This is
in contrast to other models that generally do not explain the origin of the
wisp perturbations, only their subsequent evolution. Our model naturally makes
the region between wisp 1 and wisp 2 in the optical (or the inner {\it Chandra} X-ray
ring and the torus) behave like a near zone of a radiating antenna and
prescribes the pattern of brightness fluctuations to wisps 1 and 2 and the
moving wisps. This pattern has both a stationary component oscillating in
brightness (inner {\it Chandra} ring and the torus region) and moving wisps that
cross the region. That is why the snapshots of the inner nebula taken with
insufficient temporal sampling register motion yet on average always see two
major wisps that do not seem to leave.

The morphological agreement between the behavior of the wisp region in the Crab
and the results of our simulations is an important argument in favor of the
presence of ions in the pulsar outflows, which are often presumed to consist
only of electron-positron pairs.  Such ions are a natural candidate for the
return current flow of the pulsar that starts at the auroral boundary of the
polar cap and is then mapped into the equatorial current sheet. In order to get
a fit to the nebular dynamics we require an ion injection rate of $10^{34.5}$
$\rm{s}^{-1}$.  The associated current is of the same order of magnitude as the
Goldreich-Julian current of the Crab pulsar. Thus, by studying the interaction
of the pulsar wind with the nebula we are able for the first time to get a
window on the enigmatic electrodynamics of pulsars.  Electrostatically, ions
are the preferred return current carrier for pulsars with ${\bf
\Omega} \cdot {\boldsymbol \mu} >
0$, which is thought to be the geometry applicable to the Crab pulsar
(Yadigaroglu \& Romani 1995). This means that for a rotator with
antialigned rotation and magnetic axes,
the equatorial current would be carried by electrons, and their effect on the
nebula might be different from the Crab case.

The model of the time variability of collisionless shocks presented here is
fairly general and can be applied to other PWNs, as well as other
astrophysical sources where collisionless shocks arise.  For the
pulsar B1509-58
and the associated nebula Gaensler et al. (2002) estimate a
variability period of
on the order of 3-5 yr. For
Vela the spacing of semicircular X-ray arcs implies a period of variability
similar to the Crab, yet there are no similar wisps in that remnant. This could
be explained if Vela rings are not features in the equatorial flow, but
rather in the polar direction (similar to the Crab halo (Hester et al.,
1995)). The differences might also be a consequence of the later evolutionary
state of the Vela PWN, where the environment has a chance to influence the
termination of the pulsar's wind (Chevalier 1998, 2004).
This underscores how much we are dependent on a reliable deconvolution of the
geometry and the evolutionary history in all attempts to characterize
the physics of non-thermal nebula energization by central compact
objects. For sources other than the PWNs,
collisionless shocks
are common in relativistic AGN jets and thought to occur in gamma-ray
bursts. Both settings are known to be quite variable, and whether ion-cyclotron
instability can operate in these shocks remains to be studied.

We acknowledge assistance from NASA grants
NAG5-12031 and HST-AR-09548.01-A. J.A. thanks the Miller Institute
for Basic Research in Science and the taxpayers of California for their
support. A.S. acknowledges support provided by NASA through Chandra
Fellowship grant PF2-30025 awarded by the Chandra X-Ray Center,
which is operated by the Smithsonian Astrophysical Observatory for
NASA under contract NAS8-39073.

\begin{appendix}

\section{Appendix: Model derivation and simulation details }
\label{app}
Evolution equations for a time-dependent shock in the pair-ion system can be
obtained from the divergence of the full energy-momentum tensor
$T^{\alpha \beta}=n_i m_i c^2 u_i^{\alpha} u_i^{\beta}+w u^{\alpha} u^{\beta}
+P^{\alpha \beta}+T_{\mathrm{EM}}^{\alpha \beta}$. Here $n_i m_i c^2$ is the
rest energy density for ions, ${\boldsymbol{u}}=(\gamma,
\gamma {\boldsymbol \beta})$ and
${\boldsymbol u}_i=(\gamma_i, \gamma_i {\boldsymbol \beta}_i)$ are
4-velocities of the pair and ion components, respectively, $w$ is the
enthalpy of the pairs, $P^{\alpha \beta}$ is the pressure tensor, and
$T_{\mathrm{EM}}^{\alpha \beta}$
is the electromagnetic stress-energy tensor.
   We assume that ion fluid is cold compared to pairs; thus,
only pair pressure enters into $P^{\alpha \beta}$.
Beyond the shock the
dynamics of the pairs is initially constrained to be in the plane orthogonal
to the toroidal magnetic field $B_\phi$;
therefore, we should allow for potentially
anisotropic pressure: $P^{tt}=-P_\bot$, $P^{rr}=P^{\theta \theta}=P_\bot$,
$P^{\phi \phi}=P_\parallel$.
Allowing only for  spatial variation in $r$ for all quantities,
the continuity equations together with the momentum and energy conservation
laws and induction equation
in spherical coordinates around the equatorial plane are then:

\begin{eqnarray}
&&{\partial \over \partial c t}(n_{\pm} \gamma)+{1\over r^2} {\partial \over
\partial r}(r^2 n_{\pm} \gamma \beta_r)=0, \label{pcontinuity0}\\
&&{\partial \over \partial c t}(n_{i} \gamma)+{1\over r^2} {\partial \over
\partial r}(r^2 n_{i} \gamma \beta_{ir})=0, \label{icontinuity0}\\
&&{\partial \over \partial c t}(n_i \gamma_i m_i c^2 u_{ir} + w \gamma u_r+
{1\over 4 \pi} E_\theta  B_\phi)+
{1\over r^2} {\partial \over \partial r}[r^2
(n_i m_i c^2 (u_{ir})^2+w (u_r)^2
-{1\over 8 \pi}(E_r^2 \nonumber \\  
&& \quad -E_\theta^2 - B_\phi^2))]
+{1\over r^2} {\partial \over \partial r}P_\bot+{P_\bot
-P_\parallel\over r}-{1\over r} (n_i m_i (u_{i\theta})^2+w
(u_{\theta})^2)-{E_r^2\over 4 \pi r}=0, \label{rmomentum0}\\
&&{\partial \over \partial c t}(n_i \gamma_i m_i c^2 u_{i\theta}
+ w \gamma u_{\theta}-{1\over 4 \pi} E_r  B_\phi) +
{1 \over r^2}  {\partial \over \partial r}[r^2
(n_i m_i c^2 u_{ir} u_{i\theta}+w u_r u_{\theta}-{1\over 4 \pi} E_r E_\theta)]
\nonumber \\
&& \quad +{1\over r}[n_i m_i c^2 u_{ir} u_{i\theta}+w u_r u_{\theta}-{1\over 4 \pi}
E_r E_\theta]=0, \label{thmomentum0}\\
&&{\partial \over \partial c t}[n_i m_i c^2 \gamma_i^2 + w \gamma^2
-P_\bot+{1\over 8 \pi}(E^2+B^2)]  +
{1\over r^2} {\partial \over \partial r}[r^2(n_i m_i c^2 \gamma_i
u_{ir}+ w \gamma u_r +{1\over 4 \pi} E_{\theta} B_{\phi})]=0,
    \label{energy0} \\
&& {\partial \over \partial c t} B_{\phi} + {1\over r^2} {\partial \over
\partial r}(r^2 B_\phi \beta_r)-{B_\phi \beta_r \over r}=0.
\label{induction0}
\end{eqnarray}
In (\ref{induction0}) we used the MHD condition
${\boldsymbol E}+ {\boldsymbol \beta}\times {\boldsymbol B} = 0$.
It can be shown that for variations on the ion cyclotron time-scale and
moderate harmonics thereof, departures from ideal MHD in the pairs
are negligible.

We proceed to normalize all dimensional quantities in these equations in terms
of the upstream ion gyration radius and cyclotron time as in \S\ref{model}.
Using the normalized quantities and the generalized adiabatic index
for anisotropic pressure
as defined in \S\ref{model}, the enthalpy can be written as
$w=\rho (1+\Gamma/(\Gamma-1) P_{\bot}/\rho)\equiv \rho h$
, where
$\rho$
is the rest energy density of the pairs, and $h$ is the specific
enthalpy. Using the equations
of motion for the ions,
\begin{eqnarray}
&&{d u_{ir} \over d c t}-{1\over r}{(u_{i\theta})^2\over \gamma_i}={Z
e \over m_i} [{\boldsymbol E}
+{{{\boldsymbol u}_i \over \gamma_i} \times {\boldsymbol B}}]_r \label{ionr}\\
&&{d u_{i\theta} \over d c t}+{1\over r}{u_{i\theta} u_{ir} \over
\gamma_i}={Z e \over m_i}
    [{\boldsymbol E}+{{{\boldsymbol u}_i \over \gamma_i} \times
{\boldsymbol B}}]_{\theta}, \label{iont}
\end{eqnarray}
we can rewrite the normalized pair equations
(\ref{pcontinuity0}-\ref{induction0}) as follows:
\begin{eqnarray}
&&{\partial \over \partial t}(\rho \gamma)+{1\over r^2} {\partial \over
\partial r}(r^2 \rho \gamma \beta_r)=0, \label{pcontinuity3}\\
&&{\partial \over \partial t}(\rho h \gamma^2 \beta_r+
\sigma_{\pm}\beta_r B_\phi^2)+
{1\over r^2} {\partial \over \partial r}[r^2
(\rho h \gamma^2 \beta_r^2+{1\over 2}\sigma_{\pm}(1-\beta_\theta^2+\beta_r^2)
B_\phi^2)]  \nonumber\\
&&\quad +{1\over r^2} {\partial \over \partial r}P_\bot+{P_\bot
-P_\parallel\over r}-{1\over r} \rho h \gamma^2 \beta_\theta^2
-\sigma_{\pm}{1\over r}\beta_\theta^2 B_\phi^2+\epsilon N_i (\beta_{i\theta}
-\beta_{\theta})B_\phi=0, \label{rmomentum3}\\
&&{\partial \over \partial t}(\rho h \gamma^2 \beta_{\theta}+
\sigma_{\pm} \beta_\theta B_\phi^2)+
{1 \over r^2}  {\partial \over \partial r}[r^2
(\rho h \gamma^2 \beta_r \beta_{\theta} + \sigma_{\pm} \beta_r
\beta_{\theta} B_\phi^2)]
\nonumber \\
&& \quad +{1\over r}[\rho h \gamma^2 \beta_r \beta_{\theta}+
    \sigma_{\pm} \beta_r \beta_{\theta} B_\phi^2]+\epsilon N_i
(\beta_r - \beta_{i r})B_\phi
=0, \label{thmomentum3}\\
&& {\partial \over \partial t}[\rho h \gamma^2
-P_\bot+{1\over 2} \sigma_{\pm} (1+\beta_r^2+\beta_\theta^2)B_\phi^2]
\nonumber \\
&& \quad +{1\over r^2} {\partial \over \partial r}[r^2(
    \rho h \gamma^2 \beta_r +\sigma_{\pm} \beta_r B_{\phi}^2)]
+\epsilon N_i (\beta_{i \theta} \beta_r -\beta_{i r} \beta_\theta)B_\phi=0,
    \label{energy3} \\
&&{\partial \over \partial t} B_{\phi} + {1\over r^2} {\partial \over
\partial r}(r^2 B_\phi \beta_r)-{B_\phi \beta_r \over r}=0.
\label{induction3}
\end{eqnarray}
where
energy fractions $\sigma_\pm$, $\sigma_i$ and $\epsilon$ are defined in
(\ref{sigmas}). For flows with low magnetization such that $\sigma_\pm \to 0$,
$\sigma_i \to 0$, but $\epsilon$ finite, these equations reduce to
the system (\ref{pcontinuity2}-\ref{energy2}). The Poisson equation
and Ampere's law
can be written in terms of normalized quantities as
\begin{eqnarray}
\rho_\pm+\rho_i=\sigma_i {\boldsymbol \nabla} \cdot {\boldsymbol E}
\label{chderiv},   \\
{\boldsymbol j}_\pm + {\boldsymbol j}_i =\sigma_i ({\boldsymbol \nabla}\times
{\boldsymbol B}-{\partial\over \partial t} {\boldsymbol E} \label{curderiv}),
\end{eqnarray}
where the charge densities of the pair fluid ($\rho_\pm$) and the
ions ($\rho_i$)
are normalized in terms of the upstream ion density $N_{1i}$, and similarly for
the currents ${\boldsymbol j}_\pm$, ${\boldsymbol j}_i$. When
$\sigma_i$ is small,
field gradients are on scales commensurate with
$r_{i1}$ in all
directions, and time evolution is on scales $\omega_{ci1}^{-1}$, the
expressions on the right-hand side of (\ref{chderiv}, \ref{curderiv})
can be neglected and the flow is charge and current quasi-neutral.\footnote{We
are indebted to Z.-Y. Li, who first pointed out this possible
simplification to us.}

We have developed
a code that solves the full equations (\ref{pcontinuity3}-\ref{induction3});
however,
for the region inside of the X-ray torus in the Crab Nebula, which is the
region of interest for the wisp dynamics, the inclusion of the magnetization
does not change our results. It is important to include these terms to study
the evolution of the flow in the outer nebula where the magnetic field is
compressed to equipartition, so in the following we describe the
structure of the
full simulation.

We discretize eqs. (\ref{pcontinuity2}-\ref{energy2}) on
an equidistant grid in radius ranging from $0.1$ to $10$$r_{Li1}$ with
1000 cells. Our simulation domain contains both the regions of supersonic
flow upstream of the shock and the subsonic downstream region, while
the position of the shock is determined dynamically.
To advance the pair equations, we use the Lax-Friedrichs
central-difference forward-in-time scheme. Despite its simplicity, it handles
well both the upstream and the downstream regions, including wave propagation
and scattering off the shock. To maintain
numerical stability of a centered scheme, the method introduces
inherent diffusivity, which spreads the reverse shock over 10 cells. This
is not a concern since the shock is still very well localized on the grid,
and the wave amplitude diffusion is less than 10\% over the simulation domain.
At every timestep we advance the following conservative variables:
$D=\gamma \rho$, $F=\rho h \gamma^2 \beta_r + \sigma_\pm \beta_r B_\phi^2$,
$G=\rho h \gamma^2 \beta_\theta + \sigma_\pm \beta_\theta B_\phi^2$, and
$\tau=\rho h \gamma^2 -P_\perp+{1\over 2} \sigma_\pm
(1+\beta_r^2+\beta_\theta^2) B_\phi^2$. These variables need to be
converted to the primitive
variables $\rho$, $P_\bot$, $\beta_{r,\theta}$, and $B_\phi$. We use a numerical
root-finding routine to solve for pressure $P_\bot$ from
$P_\bot=(\Gamma-1) \varepsilon_*$, where the energy density $\varepsilon_*$
is given by
\begin{equation}
\varepsilon_*=[\tau+D(1-\gamma_*)+P_\bot (1-\gamma_*^2)-\sigma_\pm B_\phi^2(2-
{1\over \gamma_*})]{1\over \gamma_*^2}.
\end{equation}

At every iteration of the root finding the Lorentz factor $\gamma_*$ is
expressed as a function of $\tau$, $F$, $G$, $P_\bot$, and $B_\phi$ through
\begin{equation}
	\gamma_*^2={(\tau+P_\bot+{1\over 2} \sigma_\pm {B_\phi^2 /
\gamma_*^2})^2 \over
		(\tau+P_\bot+{1\over 2} \sigma_\pm {B_\phi^2 /
\gamma_*^2})^2-(F^2+G^2)},
\end{equation}
which can be reduced to a cubic in $\gamma_*^2$. Having found $P_\bot$ and
$\gamma$, pair flow velocities are obtained from
\begin{eqnarray}
\beta_r={F\over \tau+P_\bot+{1\over 2} \sigma_\pm {B_\phi^2\over \gamma^2}}, \\
\beta_\theta={G\over \tau+P_\bot+{1\over 2} \sigma_\pm {B_\phi^2\over
\gamma^2}}.
\end{eqnarray}

The initial state for the calculation is the steady state strong MHD
shock. We construct this state by first computing the jump conditions
between the upstream (subscript 1) and downstream (subscript 2) side
of the shock from:
\begin{eqnarray}
\rho_1 \gamma_1 \beta_{r1} &=& \rho_2 \gamma_2 \beta_{r2}, \\
\rho_1 \gamma_1^2 \beta_{r1}^2 + {1\over 2}\sigma_\pm
(1+\beta_{r1}^2) B_{\phi1}^2 &=& \rho_2 h_2 \gamma_2^2
\beta_{r2}^2+{1\over 2} \sigma_{\pm} (1+\beta_{r2}^2)
B_{\phi2}^2+P_{\bot 2}, \\
\rho_1 \gamma_1^2 \beta_{r1}+\sigma_{\pm} \beta_{r1} B_{\phi 1}^2 &=&
\rho_2 h_2 \gamma_2^2 \beta_{r2} + \sigma_{\pm} \beta_{r2} B_{\phi 2}^2, \\
B_{\phi 1} \beta_{r 1}&=&B_{\phi 2} \beta_{r 2},
\end{eqnarray}
where the normalized quantities on the upstream side are given by
$\gamma_1=10^6$, $B_{\phi1}=1$,
$\rho_1={1\over \gamma_1^2}$, the flow is cold, $P_{\bot1}=0$, and is assumed
to have no $\theta$-component of velocity. Although there exist analytical
approximations to the jump conditions for various limiting values
of $\sigma_\pm$ (Kennel and Coroniti 1984a), we solve the general case
numerically. Having found the jump conditions, we select the radius of the
shock $r_s$ (typically, $r_s=1 r_{Li1}$) and solve for the steady state
flow from conservation laws (\ref{pcontinuity3}-\ref{energy3}) with
the time derivatives omitted.

The flow solution can be expressed in terms of constants $c_1= r^2
\gamma \rho \beta_r$ from the equation of continuity,
$c_2=r^2 \rho h \gamma^2 \beta_r^2 + r^2 \sigma_\pm \beta_r B_\phi^2$ from
the energy equation
and $c_3=r B_\phi \beta_r$ from the induction equation. These constants can
be evaluated at the shock and then used together with the equation
of state to express $P_\bot$, $\rho$, and $B_\phi$ as a function of $\beta_r$.
When substituted into the momentum equation
\begin{equation}
{1\over r^2} {\partial \over \partial r}[r^2
(\rho h \gamma^2 \beta_r^2+{1\over 2}\sigma_{\pm}(1-\beta_\theta^2+\beta_r^2)
B_\phi^2)]
+{1\over r^2} {\partial \over \partial r} P_\bot+{P_\bot
-P_\parallel\over r}=0,
\end{equation}
we get a differential equation for $\beta_r$, which is integrated numerically
inward and outward from the shock to obtain the flow solution.

The dynamical evolution equations (\ref{pcontinuity3}-\ref{energy3})
are driven by the currents induced in the pairs by the motion of the ions.
To compute the ion quantities $N_i$ and $\beta_{ir,\theta}$, we use the
1.5-dimensional particle-in-cell method. Ions are represented
as macroparticles whose velocities are calculated by advancing eqs.
(\ref{ionr}-\ref{iont}). The fields are extrapolated from the grid
to the positions of the particles, and after advance the particle
densities and currents are deposited at gridpoints. In both of these
operations we utilize volume weighting to properly account for
the diverging spherical geometry.

The simulation domain is augmented by a buffer zone of width equal to 1
$r_{Li1}$ where the field is kept constant. This zone is needed to accurately
represent the contribution of ions that would normally leave the domain and
then reenter the outer edge of the grid as part of their ${\bf E}\times {\bf
B}$ trajectory. The outer boundary conditions are designed to mimic the rest of
the nebula that is not simulated and to reduce the unwanted reflections. Two
methods are used for this. For simulations with $f \equiv r_{Li1}/r_s \ll 1$,
or essentially planar geometry, the reflections off the right edge in subsonic
outflow can be minimized by applying the method of characteristic decomposition
(Thompson 1987). The method works by explicitly setting the left-going
characteristics to zero at the boundary. For simulations with significant
spherical geometry effects ($f\sim 1$) less complicated boundary conditions of
variable extrapolation produce acceptable results.

\end{appendix}

\end{document}